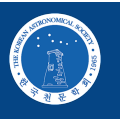

# The Outbursting YSOs Catalogue (OYCAT)

**Carlos Contreras Peña** [1,2,*], **Jeong-Eun Lee** [1], **Gregory Herczeg** [3,4], **Doug Johnstone** [5,6], **Péter Ábrahám** [7,8,9], **Simone Antoniucci** [10], **Marc Audard** [11], **Mizna Ashraf** [12], **Giseon Baek** [1,2], **Alessio Caratti o Garatti** [13], **Adolfo Carvalho** [14], **Lucas Cieza** [15,16], **Fernando Cruz-Saénz de Miera** [17,7], **Jochen Eislöffel** [18], **Dirk Froebrich** [19], **Teresa Giannini** [10], **Joel Green** [20], **Arpan Ghosh** [21], **Zhen Guo** [22,23,24], **Lynne Hillenbrand** [14], **Klaus Hodapp** [25], **Hyunwook Jheonn** [1], **Jessy Jose** [12], **Young-Jun Kim** [1], **Ágnes Kóspál** [7,8,26], **Ho-Gyu Lee** [27], **Philip W. Lucas** [28], **Tigran Magakian** [29], **Zsófia Nagy** [7], **Tim Naylor** [30], **Joe P. Ninan** [31], **S. Peneva** [32], **Bo Reipurth** [25,33], **Alexander Scholz** [34], **E. Semkov** [32], **Aurora Sicilia-Aguilar** [35], **Koshvendra Singh** [31], **Michal Siwak** [36], **Bringfried Stecklum** [18], **Zsófia Marianna Szabó** [37,34,7], **Verena Wolf** [18], and **Sung-Yong Yoon** [27]

[1] Department of Physics and Astronomy, Seoul National University, Seoul 08826, Republic of Korea
[2] Research Institute of Basic Sciences, Seoul National University, Seoul 08826, Republic of Korea
[3] Kavli Institute for Astronomy and Astrophysics, Peking University, 100871 Beijing, People's Republic of China
[4] Department of Astronomy, Peking University, 100871 Beijing, People's Republic of China
[5] NRC Herzberg Astronomy and Astrophysics, Victoria, BC, V9E 2E7, Canada
[6] Department of Physics and Astronomy, University of Victoria, Victoria, BC, V8P 5C2, Canada
[7] Konkoly Observatory, HUN-REN Research Centre for Astronomy and Earth Sciences, MTA Centre of Excellence, 1121 Budapest, Hungary
[8] Institute of Physics and Astronomy, ELTE Eötvös Loránd University, 1117 Budapest, Hungary
[9] Institute for Astronomy (IfA), University of Vienna, Türkenschanzstrasse 17, A-1180 Vienna, Austria
[10] INAF-Osservatorio Astronomico di Roma, 00078 Monte Porzio Catone, Italy
[11] Department of Astronomy, University of Geneva, 1290 Versoix, Switzerland
[12] Department of Physics, Indian Institute of Science Education and Research Tirupati, Andhra Pradesh 517619, India
[13] INAF-Osservatorio Astronomico di Capodimonte, Salita Moiariello 16, 80131 Napoli, Italy
[14] Department of Astronomy, MC 249-17, California Institute of Technology, Pasadena, CA 91125, USA
[15] Instituto de Estudios Astrofísicos, Facultad de Ingeniería y Ciencias, Universidad Diego Portales, 8370191 Santiago, Chile
[16] Millennium Nucleus on Young Exoplanets and their Moons (YEMS), Santiago, Chile
[17] Institut de Recherche en Astrophysique et Planétologie, Université de Toulouse, UT3-PS, OMP, CNRS, 31028 Toulouse Cedex 4, France
[18] Thüringer Landessternwarte Tautenburg, 07778 Tautenburg, Germany
[19] Centre for Astrophysics and Planetary Science, School of Physics and Astronomy, University of Kent, Canterbury CT2 7NH, UK
[20] Space Telescope Science Institute, 3700 San Martin Drive, Baltimore, MD 21218, USA
[21] Instituto de Radioastronomía y Astrofísica, Universidad Nacional Autónoma de México, Morelia 58089, Michoacán, México
[22] Instituto de Física y Astronomía, Universidad de Valparaíso, Casilla 5030, Valparaíso, Chile
[23] Núcleo Milenio de Formación Planetaria (NPF), Casilla 5030, Valparaíso, Chile
[24] Departamento de Física, Universidad Tecnicá Federico Santa María, Valparaíso, Chile
[25] Institute for Astronomy, University of Hawaii at Manoa, HI 96720, USA
[26] Max-Planck-Insitut für Astronomie, 69117 Heidelberg, Germany
[27] Korea Astronomy and Space Science Institute, Daejeon, 34055, Republic of Korea
[28] Centre for Astrophysics Research, University of Hertfordshire, College Lane, Hatfield, AL10 9AB, UK
[29] Byurakan Observatory NAS Armenia, Byurakan, Aragatsotn Province 0213, Armenia
[30] Department of Physics and Astronomy, University of Exeter, Exeter EX4 4QL, UK
[31] Department of Astronomy and Astrophysics, Tata Institute of Fundamental Research, Mumbai 400005, India
[32] Institute of Astronomy and National Astronomical Observatory, Bulgarian Academy of Sciences, 1784 Sofia, Bulgaria
[33] Planetary Science Institute, Tucson, AZ 85719, USA
[34] Scottish Universities Physics Alliance (SUPA), School of Physics and Astronomy, University of St Andrews, St Andrews, KY16 9SS, UK
[35] SUPA, School of Science and Engineering, University of Dundee, Nethergate DD1 4HN, Dundee, UK
[36] Mt. Suhora Astronomical Observatory, University of the National Education Commission, 30-084 Kraków, Poland
[37] Max-Planck-Institut für Radioastronomie, 53121 Bonn, Germany

*Corresponding Author: C. Contreras Peña, ccontreras@snu.ac.kr, cecontrep@gmail.com

# Abstract

Young stellar objects (YSOs) can display unpredictable and high-amplitude rises in brightness that can last from a few months to possibly over 100 years. These types of outbursts are explained by large changes in the mass accretion rate from the disk onto the central star. This type of variability has given support to a model of star formation (episodic accretion) where stars would spend most of their lifetimes accreting at low rates, and gain most of their mass through these short-lived accretion outbursts. The universality of episodic accretion, as well as its potential impact on stellar and planetary formation are still under debate. Improvement on the statistics of the members of the eruptive class is needed to better understand the episodic accretion phenomenon and its universality across different mass regimes and environments. In this paper we collect published information on the spectroscopic and photometric characteristics of 174 YSOs confirmed to belong to the eruptive variable class. We classify these objects into five different sub-classes (we find 49 FUor, 20 FUor-like, 16 EX Lupi-type, 81 Peculiar/V1647 Ori-like/MNors and 8 Periodic YSOs). The classification follows what has been done previously in the literature, and it is not an attempt to redefine these classes. In addition, we present a list of 18 embedded, and 6 massive YSOs, as additional categories of eruptive variable YSOs. Due to the complexity and/or faintness of these systems, it is hard to place them into the original classification scheme of this class of variable YSOs. Finally, we present a separate list of 355 candidate eruptive variable YSOs, which either lack spectroscopic information or the available spectroscopic data is not sufficient for an unambiguous classification. The online catalogue of confirmed and candidate eruptive YSOs will be maintained and updated in the future to serve as an important reference for the star formation community.

## 1. Introduction

Young stellar objects (YSOs) display sudden rises in brightness at optical and infrared wavelengths that provide direct observational evidence for episodic accretion, i.e., the model of star formation where stars gain most of their mass through short-lived episodes of enhanced accretion (see, e.g. Hartmann & Kenyon 1996). These YSO outbursts are usually divided into classes according to the outburst duration, strength, and spectroscopic characteristics during outburst (Audard et al. 2014; Fischer et al. 2023), with the original classification scheme arising from observations at optical wavelengths (Herbig 1977). Traditionally, the outbursts were classified into FU Ori-type (FUors, typically brighten by 4–5 magnitudes, with outburst duration of 10 years or longer, and their spectrum dominated by absorption lines) and EX Lupi-type[1] outbursts (repetitive, 2–5 magnitude, outbursts that last between a few weeks to up to 1 year, and spectrum dominated by emission lines). Observations carried out over the last decade have blurred the original classification scheme. In addition, the distinction between classes, in terms of physics of accretion and whether there is an evolutionary aspect, is not clear (Fischer et al. 2023).

The potential impact of episodic accretion on the formation of a young stellar system has gained attention over the last few decades. For example, if most of the mass is gained through short-lived episodes of high accretion, then the long periods spent at quiescence would allow the disk to cool sufficiently to fragment, helping to produce low-mass companions (Stamatellos et al. 2012). The YSO outbursts can have a long-lasting impact on the properties of the central star, such as luminosity and radius (Hartmann et al. 1997; Baraffe et al. 2017). Star formation theories still need to explain the observed luminosity spread of protostars (also known as the "luminosity problem", Kenyon et al. 1990; Evans et al. 2009; Dunham et al. 2013; Fischer et al. 2023). Episodic accretion, which plays at least a partial role, may be the most important solution.

The YSO outbursts may contribute to the build-up of the crystalline dust component ubiquitously seen in comets (Ábrahám et al. 2019; Kóspál et al. 2023). They can also alter the chemistry of protoplanetary disks and envelopes (Jørgensen et al. 2013; Molyarova et al. 2018; van't Hoff et al. 2022; Lee et al. 2023; Zwicky et al. 2024; Cruz-Sáenz de Miera et al. 2025; Lee et al. 2025) and the location of the snowline of various ices (Cieza et al. 2016). For example, the outburst in V883 Ori allowed for the identification of complex organic molecules (van 't Hoff et al. 2018; Lee et al. 2019; Jeong et al. 2025), imaging of the water snow-line (Cieza et al. 2016), and the first measurement of the D/H ratio (Tobin et al. 2023) in protoplanetary disks.

The intense accretion-related outbursts could affect the orbital evolution of planets (Boss 2013; Becker et al. 2021), the evolution of primordial dusty rings (Kadam et al. 2022) and the composition of planets (Becker & Batygin 2025). The observations of calcium–aluminum-rich inclusions in chondrites (Wurm & Haack 2009; Li et al. 2023), the depletion of lithophile elements (Hubbard & Ebel 2014) and refractory carbon in Earth (Lee et al. 2010; Klarmann et al. 2018) could all be evidence for past large eruptions in our own Solar system.

However, there are still many unanswered questions related to episodic accretion. The frequency, strength, and duration of outbursts during the protostellar evolution is still uncertain, although there has been improvement in statistics over the last few years (with outbursts becoming less frequent as YSOs evolve, see Scholz et al. 2013; Contreras Peña et al. 2019; Park et al. 2021b; Zakri et al. 2022; Contreras Peña et al. 2024).

---

[1] This class was known as EXors in the original classification of Herbig (1989). The name was established after the archetype of the class, EX Lupi, and to rhyme with "FUors" (Fischer et al. 2023). Throughout this work, and to follow the most recent literature on eruptive YSOs, we refer to this class as EX Lupi-type objects.

A number of theoretical models have been brought forward to explain the instabilities leading to an outburst, and it is not yet clear which physical mechanism governs the transport of angular momentum, or if several mechanisms are needed to explain the different types of outbursts. The various invoked mechanisms include thermal viscous instability (Bell & Lin 1994; Nayakshin et al. 2024), disk fragmentation (Vorobyov & Basu 2005, 2006, 2015), gravitational and magnetorotational instabilities (Zhu et al. 2009; Bae et al. 2014; Kadam et al. 2020), planet-disk interaction (Clarke et al. 1990; Clarke & Syer 1996; Lodato & Clarke 2004), extreme evaporation of planets (Nayakshin & Lodato 2012; Nayakshin et al. 2023), binary interactions (Bonnell & Bastien 1992), cloudlet capture (Dullemond et al. 2019), and stellar flybys (Reipurth 2000; Pfalzner 2008; Borchert et al. 2022; Dong et al. 2022).

It is still unclear if all, or even most, YSOs go through these episodes of high accretion. Millimetre continuum observations with ALMA have shown that FUor disks are more massive and compact than the disks of other classes of eruptive variables (Cieza et al. 2018), and those of regular Class II and Class I YSOs (Kóspál et al. 2021). It could be that the outburst itself modifies the structure of the disk. However, the results could imply that not all YSOs gain their mass through these large outbursts of episodic accretion, but that FUors are instead a type of YSOs that follow a particular path in their evolution that leads to the episodes of high accretion (Kóspál et al. 2021; Fischer et al. 2023).

Eruptive YSOs are still a relatively rare class of variable stars, which hinders our ability to answer the lingering questions regarding the episodic accretion phenomenon. However, the number of detected outbursts, and their relationship to lower amplitude variability, has been increasing with time thanks to continuous photometric monitoring of star-forming regions provided by e.g. *Gaia*, ZTF, Pan-STARRS, ASAS-SN, HOYS, VVV/VVVX, UKIDSS GPS, PGIR, WISE/NEOWISE, *Spitzer* and JCMT transient surveys (Hodgkin et al. 2021; Bellm et al. 2018; Chambers et al. 2016; Jayasinghe et al. 2018; Froebrich et al. 2018; Minniti et al. 2010; Saito et al. 2024; Lucas et al. 2008; De et al. 2020; Benjamin et al. 2003; Werner et al. 2004; Wright et al. 2010; Herczeg et al. 2017; Lee et al. 2021).

There exist previous compilations of eruptive YSOs in the literature since the original list of six objects from Herbig (1977). However, these are non-exhaustive (Audard et al. 2014), tend to focus on a particular class of eruptive YSOs (Lorenzetti et al. 2009; Connelley & Reipurth 2018) or are limited to sources with measured values of particular properties (Nayakshin et al. 2024). Since Herbig (1977) the number of eruptive YSOs had been increasing at a rate of 1–2 sources per year (see Figure 1). However, thanks to the aforementioned multi-epoch surveys, the number of YSO outbursts has tripled since 2014, the year of the Protostars and Planets VI chapter on eruptive YSOs (Audard et al. 2014) creating the need to provide an updated and comprehensive list of eruptive YSOs.

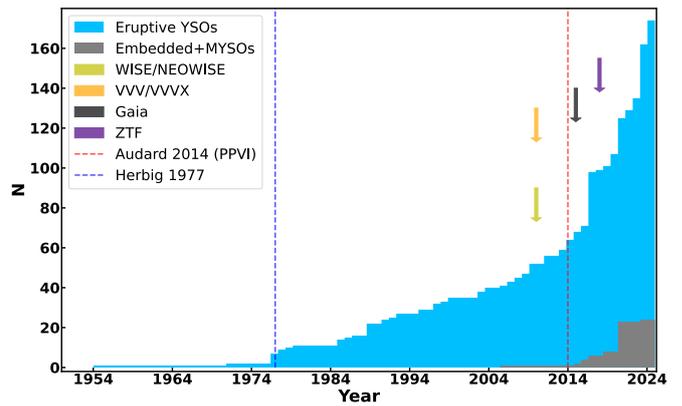

**Figure 1.** Cumulative distribution of the approximate year of discovery of eruptive variable YSOs classified in this catalogue (light blue). The same distribution is shown (grey) for the YSOs in two additional categories, embedded and MYSOs (see main text). For reference, in the figure we mark the year of publication of the catalogue of eruptive variable YSOs from Herbig (1977) (blue dashed line) and Audard et al. (2014) (red dashed line). In addition, the arrows mark the start of multi-epoch optical, near- and mid-IR surveys Gaia (black), ZTF (indigo), VVV/VVVx (orange) and WISE/NEOWISE (yellow).

The aim of this work is to provide a complete list of confirmed members of the eruptive YSO variable class. This information is provided in several tables in this paper. The same information will be provided on a website[2] that will be continuously updated as new members are confirmed.

## 2. Notes on Other Types of YSO Variability

Variability is one of the defining characteristics of YSOs (Joy 1945; Parenago 1954). Several physical mechanisms affecting the stellar photosphere, the star–disk interface, the inner edge of the dust disk, and spatial scales beyond 1 au, can drive these brightness changes (Herbst et al. 1994; Carpenter et al. 2001; Rice et al. 2015; Contreras Peña et al. 2017a; Park et al. 2021b; Lee et al. 2024). Some of the mechanisms can drive similar amplitudes and/or timescales of variability as those observed in eruptive YSOs, which can complicate the identification of sources to be included in the catalogue. Here we provide a brief discussion on the type of objects that are excluded from OYCAT.

The accretion outbursts lead to high-amplitude variability from optical to mid-IR wavelengths which can help to their identification. However, variable extinction along the line of sight can also lead to high-amplitude changes over years-long timescales, such as in YSOs AA Tau or V582 Mon (Bouvier et al. 2013; Windemuth & Herbst 2014). The eruptive activity can sometimes be combined with short-term (10–100 days) photometric fading events (typical for the UX Ori-type stars, e.g. Ábrahám et al. 2018), which can have amplitudes up to several magnitudes. The number of such examples gradually increases. However, these two types of activity probably have no connection with each other and are caused by different

---

[2] http://starformation.synology.me:5002/OYCAT/main.html

mechanisms.

The observation of stochastic (and sometimes repetitive) bursts, where accretion rates can increase by factors of 7 over 1 to 10 days, can complicate what we classify as an eruptive YSO. For example, Wang et al. (2023) shows that not every high-amplitude variation in the light curve of EX Lupi, the archetype of the EX Lupi-type class, would be classified as an EX-Lupi type outburst. The spectrum of the recent 2022 event shows photospheric $^{12}$CO bandhead absorption that is veiled due to the increase of the accretion rate by a factor of 7 (Cruz-Sáenz de Miera et al. 2023). The 2008 event, on the other hand, showed an increase in the accretion rate by a factor of 30 to 50, which led to the observation of $^{12}$CO bandhead emission typical of EX Lupi type outbursts (see Cruz-Sáenz de Miera et al. 2023, and below). In this sense, we follow the designation of Fischer et al. (2023) and classify events where accretion rates increase by factors of 10 or lower as bursts, and larger changes as outbursts. The catalogue aims to include only YSOs that have shown outbursts, and therefore, we do not include objects like TWA 3A or DQ Tau, where accretion rates periodically increase by factors of ∼3–4 (Tofflemire et al. 2017, 2025). In addition, the catalogue does not include TW Hya, a YSO with a range in accretion rates of at least a factor of 17, but these short bursts are not sustained (with timescales of ∼few days, see e.g. Herczeg et al. 2023; Wendeborn et al. 2024)

## 3. Confirmed Members

We compile a complete list of confirmed eruptive variable YSOs. The selection was done taking into account various works that have presented lists of known FUors (e.g. Audard et al. 2014; Connelley & Reipurth 2018; Szabó et al. 2023), EX Lupi-type (e.g. Lorenzetti et al. 2009, 2012; Giannini et al. 2022) and Peculiar (Contreras Peña et al. 2017b; Guo et al. 2021) outbursts. In addition we searched for works that analyse individual sources, which include outbursts from *Gaia* (e.g. Hillenbrand et al. 2018; Hodapp et al. 2019; Szegedi-Elek et al. 2020; Cruz-Sáenz de Miera et al. 2022; Nagy et al. 2023; Siwak et al. 2023; Fiorellino et al. 2024; Kuhn et al. 2024; Nagy et al. 2025; Siwak et al. 2025) and other individual discoveries (e.g. Reipurth et al. 2012; Tapia et al. 2015; Cheng et al. 2020). The current list of confirmed members of the eruptive variable class contains 164 sources, and their information is presented in Tables 2 and 3.

In Table 2, we present the main parameters of each source. This includes a source designation, right ascension, declination, YSO class and distance in columns 1 to 5. The values of extinction, $A_V$, mass accretion rate, mass of the central source, and in-outburst luminosity are presented, when available, in columns 6 to 9. In column 10 we present the year of the outburst(s) in each source. In columns 11 to 16 we show the amplitude of the variability at optical ($V, R, G$), near-IR ($K$) and mid-IR ($W1, W2$) wavelengths. As a caution to the reader, we note that the amplitudes are obtained from different sources and might not necessarily arise from contemporaneous observations. The classification of the sources from their spectra (optical to IR) during outburst, and shape of the light curve are given in columns 17 and 18 respectively. The final classification (based on the information from the previous two columns) is presented in column 19. A discussion on the complex issue of classification is presented in Section 3.1. Column 20 presents the period of the variability (if relevant to the source). Finally, the references to the information presented in the table, for each source, is presented in column 21.

Information regarding the latest available photometric observations for each source is presented in Table 3. The information is divided into optical, near-IR, mid-IR and far-IR/sub-mm wavelengths. For each wavelength range we present the Modified Julian Date (MJD), magnitude (flux), filter, and whether the observations were carried out during outburst or quiescence. A question mark denotes sources where we are not able to determine the state of the variability.

### 3.1. Classification Types

Columns 17 to 19 on Table 2 present the spectroscopic and photometric characteristics of each source, as well as the classification into the known sub-classes of eruptive variable YSOs. The eruptive variable YSOs were originally divided into two classes (FUors and EX Lupi-type) based on the distinctive characteristics displayed by a handful of objects (see Herbig 1989). Recent, multi-epoch, optical (Gaia, ZTF, Pan-STARRS, ASAS-SN), near-IR (UKIDSS GPS, VVV, PGIR), mid-IR (Spitzer, WISE, NEOWISE), and sub-mm (JCMT) surveys have led to the discovery of a larger sample of eruptive and candidate eruptive YSOs. Continuous monitoring and discoveries of eruptive variables have raised more questions about the phenomenon of accretion-related variability and have blurred the original classification system.

The issue of classification is complex and is still subject to much discussion (see e.g. Contreras Peña et al. 2023a). The aim of this work is not to redefine these categories, but rather to provide the available information for each source as taken from the works where these objects were analysed.

There are several objects that are not included in the main part of the catalogue as it is often difficult to obtain spectroscopic data to classify them, but where large and sustained accretion-related variability has been observed. These additional categories are discussed in Section 4.

In the following, we present a summary of the properties that were used to classify the eruptive YSOs in the catalogue.

**FUor**, the term FUor (coined by Ambartsumyan 1971) refers to the group of objects that share similar spectroscopic and photometric characteristics with those of the archetype of the class, FU Ori. Historical sources of this group include V1515 Cyg and V1057 Cyg (Herbig 1977). Sources in this group show an outburst that typically lasts longer than 10 years, with very unique spectra during maximum brightness. These spectral features arise from a viscously heated accretion disk that dominates emission in the system. The features include a change to later spectral type from optical to mid-IR wavelengths, $H_\alpha$ P-Cygni profile in the optical, $H_2O$ and $^{12}$CO absorption in the near-IR. For reviews and discussions on the characteristics of FUors see Herbig (1977); Hartmann & Kenyon (1996); Reipurth et al. (2010); Audard et al. (2014); Connelley & Reipurth (2018); Szabó et al. (2021); Liu et al. (2022); Rodriguez & Hillenbrand (2022); Szabó et al. (2022); Fischer et al. (2023); Carvalho et al. (2023).

Table 1. Characteristics of different classes of eruptive variable YSOs

| Class | Light Curve | Spectroscopy |
|---|---|---|
| EX Lupi-type | Short-term outbursts (duration under one year), that sometimes can be repetitive, similar to the largest outbursts in the archetype of the class EX Lupi (Kóspál et al. 2011b; Wang et al. 2023). | Dominated by emission from Hydrogen recombination lines (H$\alpha$, Br$\gamma$), metals and $^{12}$CO (e.g. Giannini et al. 2022) |
| FUor | A single outburst, where the typical timescale is longer than 10 years (Connelley & Reipurth 2018; Fischer et al. 2023) | Display all or some of the following: 1) change of spectral type from optical to near-IR wavelengths, 2) H$_\alpha$, Ca II triplet P-Cygni profile in the optical, 3) VO, TiO, H$_2$O and $^{12}$CO absorption in the near-IR, (Connelley & Reipurth 2018). |
| FUor-like | No recorded outburst | The same spectroscopic characteristics of FUors |
| Peculiar/V1647 Ori-like/MNors | The timescales of the outbursts can range from months to decades. However, the majority of the sources included in this category show outbursts which are longer than EX Lupi-type and shorter than FUors. | The sources can display emission (EX Lupi-type) or absorption (FUor) dominated spectra. This category also includes sources with spectra dominated by emission from H$_2$ lines (Guo et al. 2021; Yoon et al. 2022), or where the spectra is featureless, similar to the characteristics of e.g. OO Ser (Hodapp et al. 2012). |
| Periodic | The light curves show (quasi-) periodic variability driven by changes in the accretion rate | The sources can display emission (EX Lupi-type) or absorption (FUor) dominated spectra |

**FUor-like**, the unique spectroscopic characteristics of FUor outbursts have allowed for the classification of so-called FUor-like sources (Aspin & Reipurth 2003). These correspond to YSOs that show the same spectroscopic features as FUors, but where no outburst has been recorded (e.g. L1551-IRS5, BBW 76 and RNO54). The detection of FUor-like sources is expected because outbursts can last for decades or even centuries. Then, the initial outburst would have occurred before detailed observations and record-keeping. FUor-like sources are detected through spectroscopic studies of stars with particularly interesting properties, such as driving Herbig-Haro flows (Reipurth & Aspin 1997). However, these have been mostly serendipitous discoveries while observing samples of YSOs (e.g. Sandell & Aspin 1998).

**EX Lupi-type**, YSOs in this class (originally called EXors, after the archetype of the class EX Lupi, Herbig 1989) show outbursts that can last a few weeks to several months and reach similar amplitudes to FUor outbursts. In addition, outbursts in these systems have sometimes been seen to be recurrent (e.g. V1118 Ori Giannini et al. 2020). During the outburst, the spectra of these objects are dominated by strong emission lines, which shows that magnetospheric accretion still controls how mass is accreted onto the central star (Lorenzetti et al. 2012). In the near-IR, Na I ($\lambda$22060 Å) and $^{12}$CO bandhead ($\lambda$22935 Å) emission arise from the surface layers of a hot inner disc. These lines usually return into absorption during the quiescent state (Sipos et al. 2009; Lorenzetti et al. 2009; Rigliaco et al. 2020). During the quiescent state, the central star dominates the emission from the system, leading to the photospheric absorption profile. On the other hand, an increase in the accretion rate leads to an increase in UV radiation at the accretion shock, which in turn increases the temperature in the inner disk surface and thus favours CO emission (Lorenzetti et al. 2009). In general, EX Lupi-type objects observed at quiescent stage were found to be no different than typical T Tauri stars (Herbig 1989, 2008; Lorenzetti et al. 2012). However, some objects do differ from typical T Tauri stars (and even other EX Lupi-type objects) as they display several narrow emission metallic lines (from neutral and ionised Fe, Mg, Ca) at both outburst and quiescence (such as EX Lupi and ASASSN13db, Sicilia-Aguilar et al. 2015, 2017).

We note that there are a handful of objects that are included as they were part of the original Herbig (1977) and Herbig (1989) papers that began describing this class of eruptive YSOs. In these EX Lupi-type objects, the classification came from the observation of high-amplitude variability in photographic plates, along with changes in spectroscopic characteristics during outbursts (although this information is sparse and difficult to confirm in many cases). These changes supported variable accretion in a way that mimicked FUor outbursts but on a smaller scale (Herbig 1989). However, the available information does not allow us to confirm the change in accretion rates by factors larger than 10 and whether these outbursts were sustained (apart from EX Lupi itself). Eruptive YSOs with these issues are placed in the main catalogue but are classified as "historical" EX Lupi-type sources. These include YSOs like DR Tau, NY Ori and VY Tau.

**Objects with intermediate classifications**, many new discoveries of eruptive YSOs show spectroscopic and photometric characteristics that cannot be placed into any of the original categories described above. These new objects are often classified under different names, such as V1647 Ori-like (Fischer et al. 2023), 'peculiar' (Connelley & Reipurth 2018) or 'MNors' (Contreras Peña et al. 2017a). These objects with mixed charactersitics are divided into the categories described below

**PVM**, or a mixture of different names given to the class of YSOs with a mixture of FUor and EX Lupi-type characteristics, namely Peculiar/V1647 Ori-like/MNor. Here we include YSOs such as V1647 Ori, perhaps the archetype of eruptive YSOs with mixed characteristics, which shows multiple outbursts with duration longer than EX Lupi-type objects (i.e. 2 years or longer). The outbursts in V1647 Ori show spectroscopic characteristics of both EX Lupi-type objects and FUors (see e.g. Fischer et al. 2023). Here we also include many of the eruptive YSOs arising from the VVV survey, which are dominated by objects with emission line spectra and outbursts with duration longer than those of EX Lupi-type objects. These are interpreted as eruptive YSOs where, in spite of the large outbursts, magnetospheric accretion still controls accretion onto the central star (see Contreras Peña et al. 2017a,b; Guo et al. 2020, 2021). Similar characteristics are observed in other eruptive YSOs (e.g. ASASSN-13db, V1318 Cyg Sicilia-Aguilar et al. 2017; Hillenbrand et al. 2022).

In this category we also include eruptive YSOs that lack the FUor (absorption) or EX Lupi-type (emission) spectrum. This includes YSOs like OO Ser, a deeply embedded protostar that underwent a ∼10 years-long outburst (Hodapp et al. 1996). The spectrum at maximum brightness showed a featureless, red rising continuum (Kóspál et al. 2007). Some YSOs from the VVV survey display outflow-dominated spectra, with only $H_2$ emission, sometimes accompanied by [Fe II] emission (Guo et al. 2021). V346 Nor (Kóspál et al. 2021) also shows emission from [N II]. There is also a small group of objects that have some indication of magnetospheric accretion (Br$\gamma$ emission, but not from $^{12}CO$) and outflows ($H_2$ emission, see Guo et al. 2021).

**Periodic**, V371 Ser (EC53, Hodapp et al. 2012; Yoo et al. 2017) shows quasi-periodic variability across a wide wavelength range (1–1100 $\mu$m) with a period of $P \sim 573$ d. The variability is likely due to a cyclical buildup and draining of mass in the inner disk (Lee et al. 2020b). Periodic outbursts have also been observed in YSOs LRLL54361 (Muzerolle et al. 2013), L1634 IRS7 (Hodapp & Chini 2015), V347 Aur (Dahm & Hillenbrand 2020) and some sources from the VVV survey (Guo et al. 2021, 2022). The (quasi-) periodic nature of the variability may indicate a common physical mechanism driving the accretion changes in these systems. A stellar or planetary-mass companion interacting with the disk may induce the periodic changes (Hodapp et al. 2012; Yoo et al. 2017; Guo et al. 2022).

One object, V2492 Cyg, a YSO that went into outburst in ∼2010 and has shown quasi-periodic ($P \sim 220$ d) variability after the initial outburst, is not included in this category. The initial high-amplitude variability is driven by accretion, however, the main driver of the quasi-periodicity in the light curve are changes in the extinction along the line of sight (Hillen-brand et al. 2013). Given the reasons stated at the beginning of this section, we also do not include periodic bursters TWA3A and DQ Tau.

### 3.2. Classification of Objects in the Catalogue

Given all of the information above we present the spectroscopic and photometric information of the sources in the catalogue in the following way:

For the spectroscopic information in column **17** we add, in parenthesis, the wavelength where the spectroscopic classification arises from (O for optical and IR for near-IR). Then we provide the main spectroscopic classification as:

- **Absorption**, if the spectrum shares some or all of the spectroscopic characteristics of FUor outbursts, i.e. change of spectral type from optical to mid-IR wavelengths, $H_\alpha$ P-Cygni profile in the optical, strong water vapor and CO absorption in the near-IR (Connelley & Reipurth 2018).

- **Emission**, if the spectrum (in outburst) is dominated by emission lines and resembles the characteristics of major EX Lupi-type outbursts, i.e. emission from Hydrogen recombination lines (H$\alpha$, Br$\gamma$), Na I and $^{12}CO$ bandhead (e.g. Giannini et al. 2022)

- **Featureless**, if the spectrum lacks any of the characteristics from FUor/EX Lupi-type outbursts or outflows ($H_2$, [Fe II]), similar to the characteristics of e.g. OO Ser (Hodapp et al. 2012). If the source is featureless, but shows signatures of outflows, the classification is given as **Featureless+Outflow**. Finally, sources that in addition show some indication of accretion from Hydrogen emission lines (but lack $^{12}CO$), are classified as **Featureless+Outflow/HI**.

For the information about the light curve in column 18 we define:

- EX Lupi, this type of YSOs show short-term outbursts (duration under one year), which can be repetitive, similar to the largest outbursts in the archetype of the class EX Lupi (Kóspál et al. 2011b; Wang et al. 2023; Cruz-Sáenz de Miera et al. 2023).

- FUor, YSOs that show outbursts longer than 10 years, i.e. the typical timescales of FUor outbursts (Connelley & Reipurth 2018; Fischer et al. 2023)

- Intermediate, in these YSOs the outbursts have a duration that is longer than those expected from EX Lupi-type objects, but are shorter than the typical timescales of FUor outbursts.

- Periodic, YSOs with high-amplitude (quasi-)periodic outbursts, and where a value for the period has been determined. This category includes YSOs such as EC53 (Hodapp 1999) and V347 Aur (Dahm & Hillenbrand 2020).

- No outburst recorded, sources where no outbursts have been documented.

Based on the spectroscopic and photometric classification of each YSO, in column 19 we present the final eruptive classification of the source as either:

- EX Lupi-type, if it is classified as Emission in column 17 and EX Lupi in column 18.

- bonafide FUor, if it is classified as Absorption in column 17 and FUor in column 18.

- FUor-like, if it is classified as Absorption in column 17 and No outburst in column 18.

- Periodic, If its has spectroscopic information in column 17 and is classified as Periodic in column 18.

- PVM, if the YSOs cannot be put into any of the classes defined above.

The current version of the catalogue contains 16 EX Lupi-type (with seven of them considered as historical objects), 49 bonafide FUor, 20 FUor-like, 8 Periodic and 81 PVM sources. These classifications may change as new data are obtained in the future, and will be updated accordingly in upcoming versions of the catalogue.

## 4. Additional Categories

In addition to the discovery of new eruptive (FUor/EXlup-type) YSOs, multi-epoch and multi-wavelength surveys have allowed us to expand the investigation of episodic accretion to massive YSOs and to deeply embedded Class 0 systems (see Figure 1). Due to the complexity and/or faintness of these systems, it is sometimes hard to place the new discoveries into the original classification scheme. Therefore, we place them, for now, as separate categories that we discuss below.

### 4.1. Embedded

Surveys conducted at mid-IR, far-IR and sub-mm wavelengths have allowed the detection of outbursts in YSOs at the earlier stages of young stellar evolution (Class 0/I YSOs). These objects can be faint or invisible even at near-IR wavelengths, and lack the necessary spectroscopic information that is commonly used to classify outbursts. However, large flux changes at far-IR/sub-mm wavelengths are expected in embedded systems as a response to a change in the accretion luminosity (Johnstone et al. 2013; Balog et al. 2014).

Therefore, although not included in the main categories of this work due to a lack of spectroscopic data, we consider YSOs that showed variability at the far-IR/sub-mm as sources where variable accretion is driving the variability. Objects in this class include HOPS 383, HOPS 12 and HOPS 124 (Safron et al. 2015; Connelley & Reipurth 2020; Zakri et al. 2022) and many of the YSOs discovered by the JCMT transient survey (Lee et al. 2021; Mairs et al. 2024). Several objects that were identified as robust variables by the JCMT transient survey have spectroscopic data that supports their classification as eruptive YSOs, and are already part of the main catalogue (e.g. HOPS315, V1647 Ori). The current version of the list contains 18 embedded sources.

### 4.2. Massive YSOs

YSOs with a central mass exceeding $8\,M_\odot$ are called massive young stellar objects, or MYSOs (for a recent review, comparing low- and high-mass star formation see Beuther et al. 2025). This also includes YSOs in the earliest phases, which will form stars exceeding the above mass limit. MYSOs are rare, form in denser regions, and evolve much faster than low-mass YSOs, and feedback processes are much more important (Zinnecker & Yorke 2007).

MYSO accretion outbursts gained attention ten years ago, when two MYSO outbursts were discovered almost at the same time (S255IR–NIRS3, Caratti o Garatti et al. 2017, and references therein, and NGC 6334I–MM1, Hunter et al. 2017). These outbursts show that MYSOs can form similarly to low-mass YSOs via disk-mediated accretion. Today, episodic accretion outbursts have been confirmed with IR measurements for six MYSOs.

MYSOs are not visible in the optical, which makes them inaccessible for most ground-based surveys. Only in the later stages of their evolution (age $\geq 10^4$ yrs) they become visible in the MIR and NIR. For four of the six sources, there exist light curves in the NIR and/or MIR. For those, we provide outburst magnitudes in the Table 6. The only source for which pre- and outburst NIR spectroscopy is available is S255IR–NIRS3, and K-band spectra have been published in Caratti o Garatti et al. (2017). Outburst spectra are available for V723 Car (Tapia et al. 2015), and G323.46-0.08 (Wolf et al. 2024), in the Ks-band. All outburst spectra show emission lines, including Br$\gamma$, which comes mainly from accretion, and $H_2$, commonly associated with outflows. In S255IR–NIRS3 and V723 Car, further emission lines are present (Na I doublet, and $^{12}$CO bandhead), which in low mass YSOs, are associated with the disk as seen in EX Lupi type outbursts. While this classification seems quite reasonable for the S255IR–NIRS3 and G323.46-0.08 events (with durations of 2.5, and 8.4 yrs respectively, see Table 6), the duration of the V723 Car outburst amounts to ≈15 yrs (Tapia et al. 2015), which is quite long compared to typical EX Lupi type outbursts in low mass YSOs.

MYSOs are often associated with Class II methanol masers (microwave amplification by stimulated emission of radiation), which generally emit the strongest at 6.7 GHz (Breen et al. 2013). The high densities and temperatures of the MYSO environment are required for the excitation of this maser species. Not all MYSOs host Class II methanol masers, and there might be an evolutionary trend favoring their existence in the earliest, most embedded phases of protostellar evolution (Jones et al. 2020).

Class II methanol masers are radiatively pumped through MIR radiation (Sobolev et al. 1997). This implies, that they sensitively respond to the outburst, which makes them perfect outburst alerts. The M2O (maser monitoring organization,

Table 2. Confirmed Eruptive YSOs. Main parameters. The list contains 16 EX-Lupi-type (with seven of them considered as historical objects), 49 bonafide FUor, 20 FUor-like, 8 Periodic and 81 PVM sources.

| ID | $\alpha$ (J2000) | $\delta$ (J2000) | Class | Distance (pc) | $A_V$ (mag) | $\dot{M}$ ($M_\odot$ yr$^{-1}$) | $M_*$ ($M_\odot$) | $L$ ($L_\odot$) | Year of outburst(s)[b] | $\Delta V$ | $\Delta R$ | $\Delta G$ | $\Delta K$ | $\Delta W1$ | $\Delta W2$ | Spectroscopy ($\lambda$) | LC | Class | $P$ (d) | References |
|---|---|---|---|---|---|---|---|---|---|---|---|---|---|---|---|---|---|---|---|---|
| RNO1B | 00:36:45.99 | +63:28:52.96 | 0/II | 930 | 14.5 | $8.0 \times 10^{-6}$ | 0.2 | 1652.0 | 1978–1990 | 3.0 | — | — | — | — | — | Absorption (O+IR) | FUor | bona fide FUor | — | 25, 116, 166, 259 |
| ... | ... | ... | ... | ... | ... | ... | ... | ... | ... | ... | ... | ... | ... | ... | ... | ... | ... | ... | ... | ... |
| V733 Cep | 22:53:33.30 | +62:32:24.00 | II | 800 | 11.5 | $4.4 \times 10^{-6}$ | 0.5 | 43.0 | 1953–1984 | — | 4.5 | — | — | — | — | Absorption (O+IR) | FUor | bona fide FUor | — | 47, 62, 80, 166, 259, 276 |
| ... | ... | ... | ... | ... | ... | ... | ... | ... | ... | ... | ... | ... | ... | ... | ... | ... | ... | ... | ... | ... |
| RNO1C | 00:36:46.60 | +63:28:57.60 | II | 930 | 19.5 | $8.0 \times 10^{-6}$ | 0.2 | 1652.0 | ? | — | — | — | — | — | — | Absorption (IR) | No outburst recorded | FUor-like | — | 27, 92, 116, 166, 259 |
| ... | ... | ... | ... | ... | ... | ... | ... | ... | ... | ... | ... | ... | ... | ... | ... | ... | ... | ... | ... | ... |
| HH354IRS | 22:06:50.20 | +59:02:45.00 | 0/I | 300 | 31.5 | — | — | 16.0 | ? | — | — | — | — | — | — | Absorption (IR) | No outburst recorded | FUor-like | — | 32, 166, 259 |
| ... | ... | ... | ... | ... | ... | ... | ... | ... | ... | ... | ... | ... | ... | ... | ... | ... | ... | ... | ... | ... |
| XZ Tau[a] | 04:31:40.07 | +18:13:57.20 | II | 150 | 3.0 | $1.0 \times 10^{-6}$ | 0.4 | 10.7 | m1930?, 1990, 2014 | 6.2 | — | — | — | — | — | Emission (O+IR) | EXLupi | bona fide EX Lupi | — | 42, 44, 60, 71, 236, 259 |
| ... | ... | ... | ... | ... | ... | ... | ... | ... | ... | ... | ... | ... | ... | ... | ... | ... | ... | ... | ... | ... |
| GM Cep | 21:38:17.33 | +57:31:22.01 | II | 900 | 3.0 | $3.0 \times 10^{-7}$ | 2.1 | 40.0 | 1986, 2003, 2007 | 2.1 | 2.0 | — | — | — | — | Emission (O) | EXLupi | bona fide EX Lupi | — | 68, 96 |
| ... | ... | ... | ... | ... | ... | ... | ... | ... | ... | ... | ... | ... | ... | ... | ... | ... | ... | ... | ... | ... |
| V1180 Cas | 02:33:01.54 | +72:43:26.81 | I | 600 | 3.3 | $3.0 \times 10^{-8}$ | 0.8 | — | 2000, 2004, 2011, 2020 | 5.3 | 4.0 | — | 1.0 | — | — | Emission (O+IR) | Intermediate | PVM | — | 88, 111, 112, 241, 259 |
| ... | ... | ... | ... | ... | ... | ... | ... | ... | ... | ... | ... | ... | ... | ... | ... | ... | ... | ... | ... | ... |
| 2MASS J22352345 +7517076 | 22:35:23.46 | +75:17:07.60 | I | 350 | 9.4 | $1.1 \times 10^{-4}$ | 1.8 | 165.0 | 1993–1998 | — | — | — | 5.0 | 4.0 | 3.5 | Featureless+ Outflow/HI (O+IR) | FUor | PVM | — | 132, 178 |
| ... | ... | ... | ... | ... | ... | ... | ... | ... | ... | ... | ... | ... | ... | ... | ... | ... | ... | ... | ... | ... |
| LRLL54631 | 03:43:51.02 | +32:03:08.1 | 0 | 294 | 5.9 | $1.0 \times 10^{-6}$ | 0.2 | 2.7 | 2004p | — | — | — | — | 3.0 | 2.8 | Emission (IR) | Periodic | Periodic | 25.3 | 102, 106, 183, 272 |
| ... | ... | ... | ... | ... | ... | ... | ... | ... | ... | ... | ... | ... | ... | ... | ... | ... | ... | ... | ... | ... |
| V371 Ser | 18:29:51.20 | +01:16:39.00 | I | 429 | 47.5 | $8.0 \times 10^{-6}$ | 0.3 | 1.6 | 1994p | — | — | — | 1.5 | 1.7 | 1.9 | Absorption (IR) | Periodic | Periodic | 530.0 | 35, 161, 166, 185, 200, 201 |

[a]Objects in the EX-Lupi type class that are defined as "historical" (see main text). [b]**Year1,Year2,Year3**: Outbursts in Year1, Year2 and Year3; **<Year1**: The outburst occurred somewhere between Year1 and Year2; **Year1?**: The exact year of outburst is uncertain; **mYear1,Year2,Year3**: Many outbursts have been observed, some of them occurred in Year1, Year2 and Year3; **Year1p**: The source shows periodic outbursts, one of them was observed in Year1. The entire table is available at https://github.com/JKAS-Hub/Supp/b1ob/main/v58n2p209_Tab1es.pdf

Table 3. Confirmed Eruptive YSOs. Latest available photometry.

| ID | In Outburst? | Optical | | | | Near-IR | | | | Mid-IR | | | | Far-IR/Sub-mm | | | | References |
|---|---|---|---|---|---|---|---|---|---|---|---|---|---|---|---|---|---|---|
| | | MJD | Mag | Band | State | MJD | Mag | Band | State | MJD | Mag | Band | State | MJD | Flux (Jy) | Band (μm) | State | |
| RNO 1B | Y | 60488.3 | 19.1 | ztf r | Out. | 57262.0 | 8.3 | K | Out. | 60161.8 | 4.7 | WISE W2 | Out. | 50934.0 | 6.6 | 850 | Out. | 37, 122, 163, 166 |
| ... | ... | ... | ... | ... | ... | ... | ... | ... | ... | ... | ... | ... | ... | ... | ... | ... | ... | ... |
| V733 Cep | Y | 60490.4 | 18.0 | ztf r | Out. | 57199.0 | 8.3 | K | Out. | 60146.5 | 6.9 | WISE W2 | Out. | 56316.0 | 7.5 | 160 | Out. | 122, 163, 166, 275 |
| ... | ... | ... | ... | ... | ... | ... | ... | ... | ... | ... | ... | ... | ... | ... | ... | ... | ... | ... |
| RNO 1C | Y | 60488.4 | 19.5 | ztf r | Out. | 57262.0 | 7.7 | K | Out. | 60323.2 | 4.7 | WISE W2 | Out. | 50934.0 | 6.6 | 850 | Out. | 37, 122, 163, 166 |
| ... | ... | ... | ... | ... | ... | ... | ... | ... | ... | ... | ... | ... | ... | ... | ... | ... | ... | ... |
| HH354IRS | Y | 55735.0 | 19.4 | Pan-STARRS y | Out. | 57325.0 | 10.8 | K | Out. | 60131.2 | 8.5 | WISE W2 | Out. | 56299.0 | 8.5 | 500 | Out. | 122, 137, 166, 180 |
| XZ Tau[a] | N | 60623.4 | 14.8 | ASAS-SN g | Quiesc. | 51493.0 | 7.3 | Ks | Quiesc. | 60187.9 | 4.9 | WISE W2 | Quiesc. | 57044.0 | 0.1 | 1300 | Quiesc. | 54, 122, 169, 170 |
| ... | ... | ... | ... | ... | ... | ... | ... | ... | ... | ... | ... | ... | ... | ... | ... | ... | ... | ... |
| GM Cep | N | 60621.2 | 14.2 | ASAS-SN g | Quiesc. | 54259.0 | 8.5 | K | Quiesc. | 60489.0 | 7.0 | WISE W2 | Quiesc. | 56308.0 | 0.9 | 160 | Quiesc. | 68, 122, 169, 275 |
| V1180 Cas | Y | 60392.2 | 16.4 | ztf r | Out. | 56738.0 | 10.9 | K | Out. | 60186.2 | 8.0 | WISE W2 | Out. | 48381.0 | 8.7 | 100 | Quiesc. | 111, 112, 122, 124, 163 |
| ... | ... | ... | ... | ... | ... | ... | ... | ... | ... | ... | ... | ... | ... | ... | ... | ... | ... | ... |
| 2MASS J22352345 +7517076 | Y | — | — | — | — | 58036.0 | 9.1 | Ks | — | 60330.0 | 1.4 | WISE W2 | Out. | 51544.0 | 5.7 | 850 | Quiesc. | 65, 122, 178 |
| LRLL54631 | ? | 57000.4 | 21.2 | Pan-STARRS r | Quiesc. | 55511.0 | 15.1 | K | Quiesc. | 60339.3 | 8.0 | WISE W2 | Out. | 52680.0 | 0.5 | 1100 | Quiesc. | 59, 102, 122, 137 |
| ... | ... | ... | ... | ... | ... | ... | ... | ... | ... | ... | ... | ... | ... | ... | ... | ... | ... | ... |
| V371 Ser | Y | — | — | — | — | 58907.6 | 10.9 | K | Out. | 60396.7 | 6.8 | WISE W2 | Out. | 53551.0 | 1.5 | 1100 | Quiesc. | 102, 122, 200 |

[a] Objects in the EX-Lupi type class that are defined as "historical" (see main text). The entire table is available at https://github.com/JKAS-Hub/Supp/blob/main/v58n2p209_Tables.pdf

**Table 4.** Embedded YSOs. Main parameters.

| ID | α (J2000) | δ (J2000) | Class | Distance (pc) | $A_V$ (mag) | $\dot{M}$ ($M_\odot$ yr$^{-1}$) | $M_*$ ($M_\odot$) | $L$ ($L_\odot$) | Year of outburst(s)[a] | ΔV | ΔR | ΔG | ΔK | ΔW1 | ΔW2 | Spectroscopy (λ) | LC | Class | P (d) | References |
|---|---|---|---|---|---|---|---|---|---|---|---|---|---|---|---|---|---|---|---|---|
| NGC1333VLA3 | 03:29:03.372 | +31:16:01.60 | — | 293 | — | — | — | — | 2017? | — | — | — | — | 2.0 | 1.5 | — | FUor? | Embedded | — | 223, 274 |
| WEST40 | 03:29:04.06 | +31:14:46.5 | 0 | 293 | 5.9 | — | — | 0.7 | 2017 | — | — | — | — | 0.6 | 0.4 | — | FUor? | Embedded | — | 102, 157, 223, 274 |
| IRAS4A | 03:29:10.49 | +31:13:30.8 | 0 | 270 | 5.9 | — | — | 9.8 | 2012–2016 | — | — | — | — | — | 1.1 | — | FUor? | Embedded | — | 102, 153, 223, 265, 274 |
| … | … | … | … | … | … | … | … | … | … | … | … | … | … | … | … | … | … | … | … | … |
| SERPENS SMM10 | 18:29:52.20 | +01:15:47.6 | I | 436 | 28.6 | — | 0.5 | 7.0 | 2012–2016 | — | — | — | — | 1.4 | 1.1 | — | FUor? | Embedded | — | 223, 256, 274 |
| CARMA7 | 18:30:04.10 | −02:03:02.5 | 0 | 436 | 140.2 | — | 0.5 | 50.3 | 2012–2016? | — | — | — | — | — | — | — | FUor? | Embedded | — | 223, 256, 274 |
| B335 | 19:37:01.03 | +07:34:10.90 | 0 | 164 | — | $1.0 \times 10^{-5}$ | 0.2 | 18.0 | 2010–2013 | — | — | — | — | — | 2.4 | — | Intermediate | Embedded | — | 270 |

[a]Same as Table 2. The entire table is available at https://github.com/JKAS-Hub/Supp/blob/main/v58n2p209_Tables.pdf

**Table 5.** Embedded YSOs. Latest available photometry.

| | Optical | | | | | Near-IR | | | | | Mid-IR | | | | | Far-IR/Sub-mm | | | | |
|---|---|---|---|---|---|---|---|---|---|---|---|---|---|---|---|---|---|---|---|---|
| ID | in outburst? | MJD | Mag | Band | State | MJD | Mag | Band | State | MJD | Mag | Band | State | MJD | Flux (Jy) | Band (μm) | MJD | State | References |
| NGC1333VLA3 | Y | — | — | — | — | — | — | — | — | 60336.3 | 9.3 | WISE W2 | Out. | 3.5 | 850 | 60190.0 | Out. | 121, 273 |
| WEST40 | Y | — | — | — | — | — | — | — | — | 60336.3 | 11.8 | WISE W2 | Out. | 0.6 | 850 | 60190.0 | Out. | 121, 273 |
| IRAS4A | N | — | — | — | — | — | — | — | — | 60177.9 | 14.2 | WISE W2 | Quiesc. | 9.2 | 850 | 60190.0 | Quiesc. | 121, 273 |
| … | … | … | … | … | … | … | … | … | … | … | … | … | … | … | … | … | … | … |
| SERPENS SMM10 | ? | — | — | — | — | — | — | — | — | 60396.7 | 8.7 | WISE W2 | Quiesc. | 0.8 | 850 | 60196.0 | Quiesc. | 121, 273 |
| CARMA7 | Y | — | — | — | — | — | — | — | — | — | — | — | — | 4.8 | 850 | 60182.0 | Out. | 273 |
| B335 | N | — | — | — | — | — | — | — | — | 60417.3 | 11.2 | WISE W2 | Quiesc. | 2.2 | 850 | 51544.0 | Quiesc. | 64, 121 |

The entire table is available at https://github.com/JKAS-Hub/Supp/blob/main/v58n2p209_Tables.pdf

Table 6. Eruptive Massive YSOs

| ID | α (J2000) | δ (J2000) | ClassII maser | Distance (pc) | $A_V$ (mag) | $\dot{M}$ ($M_\odot$ yr$^{-1}$) | $M_*$ ($M_\odot$) | $L_{pre}$ ($L_\odot$) | $L_{peak}$ ($L_\odot$) | $t_{rise}$ | $\Delta t$ | $E_{acc}$ ($10^{45}$ erg) | $\dot{M}_{acc}$ ($M_{Jup}$) | $\Delta K$ | $\Delta W1$ | $\Delta W2$ | Spectroscopy (λ) | LC | Class | P (d) | References |
|---|---|---|---|---|---|---|---|---|---|---|---|---|---|---|---|---|---|---|---|---|---|
| G323.46 -0.08 | 15:29:19.59 | −56:31:21.9 | Yes | $4080^{+400}_{-380}$ | 18 ± 1 | $8 \times 10^{-4}$ | 23 | $6 \times 10^4$ | $32 \times 10^4$ | 1.4 | 8.4 | 90 | 7.3 | ≈2.5 | ≈1.5 | ≈1 | Emission (IR) | — | MYSO | — | 181, 279 |
| S255IR-NIRS3 | 06:12:54.013 | +17:59:23.05 | Yes | $1780^{+110}_{-120}$ | 44 ± 16 | $5 \times 10^{-3}$ | 20 | $2.9 \times 10^4$ | $15.9 \times 10^4$ | 0.4 | 2.5 | 12 | 2 | ≈3.4 | ≈1.3 | ≈1 | Emission (IR) | — | MYSO | — | 147, 171, 202 |
| G358.93 -0.03-MM1 | 17:43:10.02 | −29:51:45.8 | Yes | $6750^{+370}_{-680}$ | 60 ± 10 | $3.2 \times 10^{-3}$ | 12 ± 3 | $5 \times 10^3$ | $2.4 \times 10^4$ | 0.1 | 0.5 | 2.8 | 0.6 | — | — | — | — | — | MYSO | — | 174, 187, 189, 229 |
| NGC 6334I-MM1 | 17:20:53.4 | −35:46:57 | Yes | $1260^{+330}_{-210}$ | — | $2.3 \times 10^{-3}$ | 6.7 | $3 \times 10^3$ | $4.9 \times 10^4$ | 0.6 | 8 | 32 | 0.3 | — | — | — | — | — | MYSO | — | 154, 164, 173, 220 |
| V723 Car | 10:43:23.25 | −59:33:56.9 | No | 2500 ± 200 | 55 | — | 10 | — | $4 \times 10^3$ | 4 | 15 | — | — | >4.3 | — | — | Featureless/Emission (IR) | — | MYSO | — | 55, 134 |
| M17 MIR | 18:20:23.017 | −16:11:47.98 | No | 1900 ± 100 | — | $1.7 \times 10^{-3}$ | 5.4 | $1.4 \times 10^3$ | $9 \times 10^3$ | — | 20 | — | — | — | — | ≈2.1 | — | — | MYSO | — | 212, 280 |

The entire table is available at https://github.com/JKAS-Hub/Supp/blob/main/v58n2p209_Tables.pdf

Table 7. Candidate Eruptive YSOs.

| ID | α (J2000) | δ (J2000) | Class | Distance (pc) | Δ | Magnitude | MJD | Filter | Spectroscopy (λ) | LC | P (d) | References |
|---|---|---|---|---|---|---|---|---|---|---|---|---|
| YSO 2099 | 15:29:06.2 | −56:23:10 | FS | 3710 | 5.6 | 16.2 | 58704.1 | $K_s$ | — | FUor | — | 264 |
| HOPS 20 | 05:33:30.82 | −05:50:40 | I | 389 | 5.0 | 16.3 | 56954.3 | K | — | FUor | — | 208, 226 |
| L222_42 | 16:29:39.03 | −49:01:15 | II | 3000 | 4.9 | 16.1 | 58704.16 | $K_s$ | — | EX Lupi | — | 222, 273 |
| L222_55 | 16:49:31.54 | −45:07:15 | — | — | 4.8 | 13.5 | 58727.09 | $K_s$ | — | EX Lupi | — | 273 |
| Lucas et al. (2017) 282 | 19:19:50.47 | +14:03:07 | — | 10600 | 4.6 | 13.4 | 55787.3 | K | — | — | — | 156 |
| L222_84 | 17:24:36.93 | −34:08:30 | FS | — | 4.5 | 16.5 | 58723.13 | $K_s$ | — | Intermediate | — | 273 |
| Contreras Peña et al. (2019) V4 | 02:33:53.40 | +61:56:50 | II | 2100 | 4.5 | 16.5 | 57204 | R | — | FUor | — | 175, 211 |
| L222_164 | 17:49:33.19 | −26:57:08 | — | — | 4.2 | 15.2 | 58715.18 | $K_s$ | — | Intermediate | — | 273 |
| ... | ... | ... | ... | ... | ... | ... | ... | ... | ... | ... | ... | ... |

The entire table is available at https://github.com/JKAS-Hub/Supp/blob/main/v58n2p209_Tables.pdf

Burns 2024) has continuously been observing a large sample of suitable sources since 2017, which enabled the first discovery of an MIR dark accretion outburst (G358.93-0.03-MM1, Sugiyama et al. 2019).

MYSOs typically emit the bulk of their energy at far-infrared wavelengths. The increase in dust continuum emission due to the outburst represents proof of its presence and allows us to derive its energy.

In this work, we include only episodic outbursts that have been confirmed with IR data. We do not include periodic sources (e.g. G24.33+0.14, Hirota et al. 2022, among others). However, some sources such as S255IR-NIRS3 (Burns et al. 2016), and M17 MIR (Chen et al. 2021; Zhou et al. 2024) show hints for multiple outbursts.

The current version of the catalogue contains six eruptive MYSOs (see Table 6).

### 4.3. Candidate Eruptive Variable YSOs

We have also compiled a list of YSOs that have been listed as candidate eruptive variable YSOs. This list contains YSOs that are suspected to be eruptive due to the observation of a large photometric brightening towards the source. These YSOs show high-amplitude near- and/or mid-IR variability, but a) have no spectroscopic follow-up to confirm their membership, or b) have spectroscopic follow-up, but the authors in the respective works have not been able to confirm membership based on this information. The list is comprised of 355 YSOs.

The list of candidates has been compiled from several works that analyse multi-epoch near- and mid-IR photometry from the UKIDSS Galactic Plane Survey (Contreras Peña 2015; Lucas et al. 2017), the Vista Variables in the Via Lactea Survey (Contreras Peña et al. 2017a; Guo et al. 2022), and the WISE/NEOWISE surveys (Antoniucci et al. 2014b; Park et al. 2021b; Contreras Peña et al. 2023a). The list contains all of the YSOs that show $\Delta K_s > 1$ mag, and were classified as eruptive from their light curves in Contreras Peña et al. (2017a). In the case of the candidates from Lucas et al. (2017), we only selected objects that are classified as YSOs and show $\Delta K > 2$ mag in that particular work. The reason for the difference in the amplitude selection is that Contreras Peña et al. (2017b) based their selection on the inspection of light curves with at least 44 epochs of $K_s$ photometry, whilst the analysis of Lucas et al. (2017) is based on 2 to 3 epochs of K-band photometry. Therefore using a larger amplitude cut in the sample of the latter work will likely introduce a lower number of false positives into the sample of candidate eruptive variable YSOs.

Table 7 contains the information for the list of candidates. The column information is similar to that of Tables 2 and 3. The only differences are that we only present one value for the amplitude in column 6, and specify the filter in which this amplitude was estimated in column 7. Information for the latest available observation at the specified filter is given in columns 8 and 9, respectively.

## 5. Initial Comparison of Parameters

The parameters presented in Table 2 allow us to search if the sub-classes of eruptive variable YSOs show any particular trends that might be useful in a future re-definition of the classification scheme. The comparison is useful as theoretical models show that parameters like the ntmass of the central star and the maximum accretion rate during outburst can lead to different spectroscopic properties of the YSO (Calvet et al. 1991; Liu et al. 2022; Rodriguez & Hillenbrand 2022). Any analysis from these parameters is limited as there are several uncertainties in these measurements. In addition, many sources lack appropriate measurements. Finally, it is not the aim of this work to re-define the classification scheme.

In Figure 2 we show the distribution of YSO class, distance, extinction, accretion rate, stellar mass and in-outburst luminosity for sources in each class of eruptive variable YSOs. We test for differences in the distributions using a Mann-Whitney U test. The sample sizes are not large for some of the parameters, so any conclusions may be affected by low-number statistics. This is especially true when comparing with Periodic sources, as there are only eight YSOs in this class.

The PVM, bonafide FUors and FUor-like sources tend to have the largest accretion rates and luminosities compared with EX Lupi-type and periodic sources. Bonafide FUors and PVM sources are also distributed toward larger values of distance. The comparison shows bonafide FUors are, on average, more distant than FUor-like sources. However, there are no notable differences between the two classes when comparing the distribution of accretion luminosities, mass, extinction, class and accretion rates. The large similarity between bonafide FUors and FUor-like YSOs is not surprising, as the only criterion that separates the two is the lack of a recorded outburst in the latter group. EX Lupi-type objects show the lowest accretion rates, extinction and luminosities compared with PVM, bonafide FUors and FUor-like sources. The EX Lupi-type outbursts are also found predominantly at older evolutionary stages than these three groups. Eruptive YSOs in the PVM class also show larger masses compared with bonafide FUors, FUor-like and periodic sources. We find no differences when comparing with the mass distribution of EX Lupi-type sources.

## 6. Summary

In this work, we provide a comprehensive list of confirmed and candidate eruptive variable YSOs. We present, when available, information on stellar parameters, accretion rate, luminosity, and photometric observations, among others, for 174 YSOs that are confirmed to be eruptive YSOs. Similar, but more limited information, is provided for 355 YSOs that are classified as candidate eruptive variable YSOs.

Using the available photometric and spectroscopic information of confirmed sources, we classify these objects into five different sub-classes (FUor, FUor-like, EX Lupi-type, Peculiar/V1647 Ori-like/MNors and Periodic). The classification follows what has been done previously in the literature, and it is not an attempt to redefine these classes.

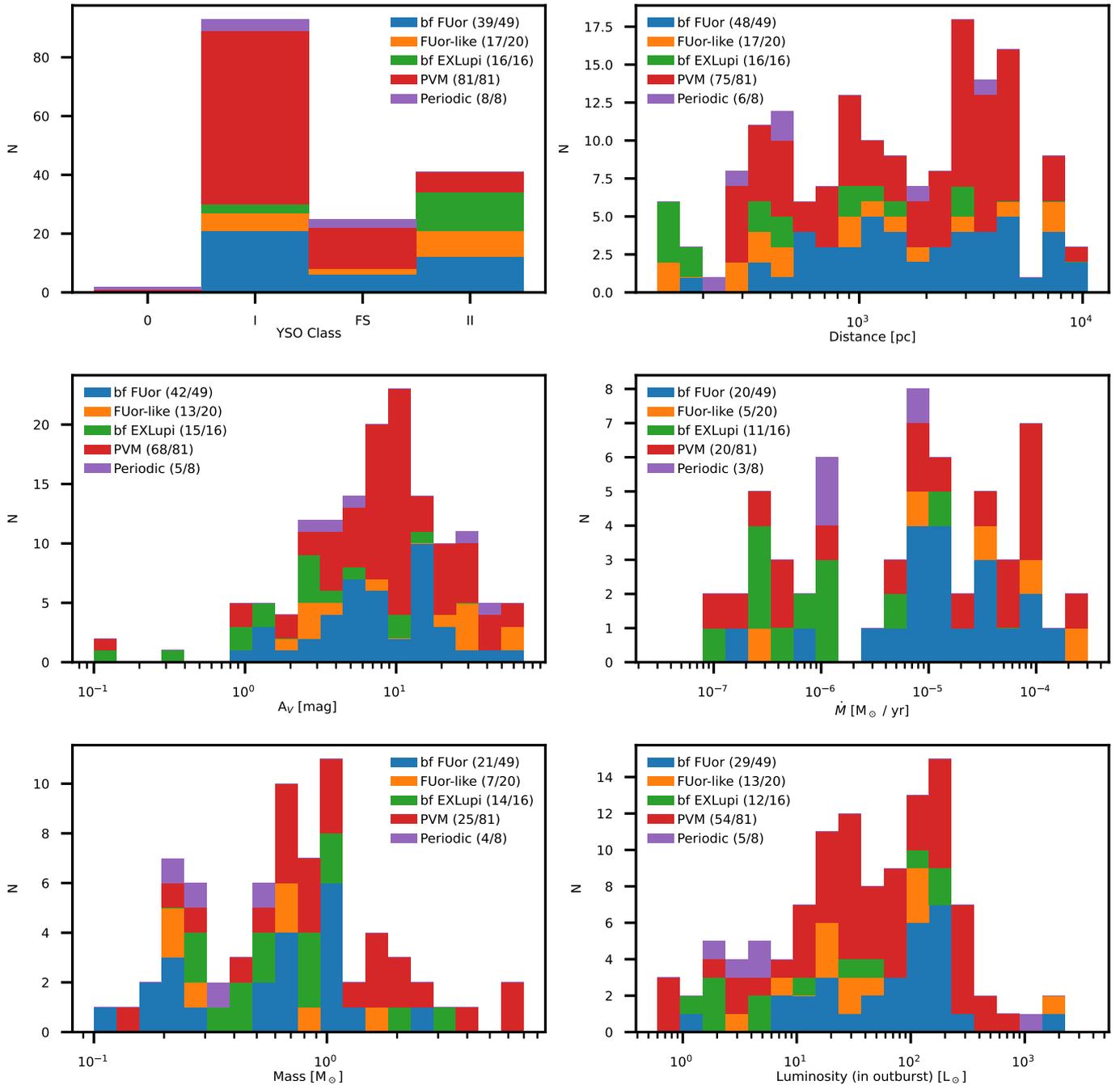

**Figure 2.** Distribution of parameters presented in Table 2 divide for each sub-class of confirmed eruptive variable YSOs. Next to each label we show the number of sources in each class that have estimates for the specific parameter.

The catalogue also includes YSOs where large and sustained accretion-related outbursts have been observed, but it is often difficult to obtain spectroscopic data to place them into the known sub-classes of eruptive variable YSOs. These additional categories include outburts in massive YSOs where the mass of the central star exceeds 8 $M_\odot$ (MYSOs) and in YSOs that are often too faint or invisible at optical to near-IR wavelengths (Embedded).

An initial comparison of parameters (such as central mass, distance, luminosity) from the different subclasses reveals some differences. For example EX Lupi-type sources tend to be more evolved and have the lower in-outburst accretion luminosities compared with other classes. PVM sources tend, on average, to be more massive than other YSOs in the catalogue. However, the comparison suffers from many caveats, such as low number statistics, lack of measured parameters in many sources, and/or differences on how the values of the same parameter were estimated.

The purpose of the catalog is to be a useful tool for the community. The online version will be continuously updated as new sources are discovered.

## Acknowledgments

This publication makes use of data products from the Near-Earth Object Wide-field Infrared Survey Explorer (NEO-WISE), which is a joint project of the Jet Propulsion Laboratory/California Institute of Technology and the University of California, Los Angeles. NEOWISE is funded by the National Aeronautics and Space Administration

C.C.P. is grateful to the scientific and local organizing committees of the "Born in Fire: Eruptive Stars and Planet Formation" conference that was held on September 24–27, 2024 at Universidad Diego Portales, Santiago, Chile. The discussions held at the conference were very important in completing and improving this catalog.

C.C.P. was supported by the National Research Foundation of Korea (NRF) grant funded by the Korean government (MEST) (No. 2019R1A6A1A10073437). J.E.L. was supported by the New Faculty Startup Fund from Seoul National University and the NRF grant funded by the Korean government (MSIT) (grant number 2021R1A2C1011718 and RS-2024-00416859). GJH is supported by the National Key R&D program of China 2022YFA1603102 and by general grant 12173003 from the National Natural Science Foundation of China. D.J. is supported by NRC Canada and by an NSERC Discovery Grant. A.K. was supported by the NKFIH excellence grant TKP2021-NKTA-64. Zs.N. acknowledges the Hungarian National Research, Development and Innovation Office grant OTKA FK 146023. Zs.N. was supported by the János Bolyai Research Scholarship of the Hungarian Academy of Sciences. Zs.N. acknowledges support from the ESA PRODEX contract No. 4000132054. Zs.M.Sz. acknowledges funding from a St. Leonards scholarship from the University of St. Andrews. Zs.M.Sz. is a member of the International Max Planck Research School (IMPRS) for Astronomy and Astrophysics at the Universities of Bonn and Cologne. A.C.G. acknowledges support from PRIN-MUR 2022 20228JPA3A "The path to star and planet formation in the JWST era (PATH)" funded by NextGeneration EU and by INAF-GoG 2022 "NIR-dark Accretion Outbursts in Massive Young stellar objects (NAOMY)" and Large Grant INAF 2022 "YSOs Outflows, Disks and Accretion: towards a global framework for the evolution of planet forming systems (YODA)". F.C.S.M. received financial support from the European Research Council (ERC) under the European Union's Horizon 2020 research and innovation programme (ERC Starting Grant "Chemtrip", grant agreement No 949278). G.B was supported by Basic Science Research Program through the National Research Foundation of Korea(NRF) funded by the Ministry of Education(grant number RS-2023-00247790). Z.G. is funded by ANID, Millennium Science Initiative, AIM23-001. Z.G. is supported by the China-Chile Joint Research Fund (CCJRF No.2301) and Chinese Academy of Science South America Center for Astronomy (CASSACA) Key Research Project E52H540301. This work was also supported by the NKFIH NKKP grant ADVANCED 149943. Project no.149943 has been implemented with the support provided by the Ministry of Culture and Innovation of Hungary from the National Research, Development and Innovation Fund, financed under the NKKP ADVANCED funding scheme.

# Appendix A. Photometry

The information provided in Tables 3 and 5 arises from extensive search through past publications and the VIZIER database. In addition we collected data from several surveys we introduce below.

## Appendix A.1. Gaia

For each source we collected, when available, data for the G, GBP and GRP passbands from the Gaia DR3 database (Gaia Collaboration et al. 2023). In addition, we collect G-band light curves of sources that displayed high-amplitude variability and were detected as part of the *Gaia* alerts database (Hodgkin et al. 2021). The Gaia DR3 covers observations between July 2014 and May 2017, while the Gaia alerts database provides observations between July 2014 and January 2025.

## Appendix A.2. PanSTARRS/ZTF

For YSOs located at declination north of $-30$ deg, we collected optical $(g, r, i)$ light curves from the Panoramic Survey Telescope and Rapid Response System (Pan-STARRS, Chambers et al. 2016) data release DR2 using the Mikulski Archive for Space Telescopes (MAST). In addition we collected data from the Zwicky Transient Facility (ZTF, Bellm et al. 2019) data release DR22 using the InfraRed Science Archive(IRSA). ZTF DR22 contains observations from 2018 until February 2025.

## Appendix A.3. WISE

This work uses mid-IR photometry from all-sky observations of the *WISE* telescope. *WISE* surveyed the entire sky in four bands, W1 (3.4 $\mu$m), W2 (4.6 $\mu$m), W3 (12 $\mu$m), and W4 (22 $\mu$m), with the spatial resolutions of 6.1 arcsec, 6.4 arcsec, 6.5 arcsec, and 12 arcsec, respectively, from 2010 January to September (Wright et al. 2010). The survey continued as the NEOWISE Post-Cryogenic Mission, using only the W1 and W2 bands, for an additional 4 months (Mainzer et al. 2011). In September 2013, WISE was reactivated as the NEOWISE-reactivation mission (NEOWISE-R, Mainzer et al. 2014). NEOWISE-R continued to operate until the decomissioning of the WISE telescope in July 2024. For each visit to a particular area of the sky, *WISE* performs several photometric observations over a period of ∼few days. Each area of the sky is observed similarly every ∼6 months.

For the analysis of eruptive YSOs we used all the available data from the *WISE* telescope for observations from 2010 until July 2024. The single-epoch data was collected from the NASA/IPAC Infrared Science Archive (IRSA) catalogues using a 3 arcsec radius from the coordinates of the YSO. For each source, we averaged the single epoch data taken a few days apart to produce 1 epoch of photometry every 6 months (following the procedures described in Park et al. 2021b).

## Appendix A.4. JCMT Transient/GBS Maps

The JCMT Transient Survey uses the SCUBA-2 instrument (Holland et al. 2013) on JCMT to monitor sub-mm contin-

uum emission from eight nearby star-forming regions (Ophiuchus Core, NGC 1333, IC 348, Serpens Main, Serpens South, OMC 2/3, NGC 2024, and NGC 2068 Herczeg et al. 2017; Mairs et al. 2024). Each region is observed in a PONG mode that produces an image with smooth sensitivity across a field with 30 arcmin diameter, with an integration time set to reach $\sim$12 mJy at 850 $\mu$m. The data are reduced using customized routines, including spatial masks and offsets, with the map-making software, MAKEMAP (see Chapin et al. 2013, , for details) in the STARLINK package (Jenness et al. 2013; Currie et al. 2014). The 850 $\mu$m fluxes are measured from the peak brightness of the object, and are then calibrated using bright sources that are measured to be non-varying at 850 $\mu$m. A full description of our reductions and calibrations is described by (Mairs et al. 2017a; Mairs et al. 2024) . The flux calibration uncertainty in any single epoch is $\sim 0.025 \times$ F850+12 [mJy], determined by a combination of the noise level and the stability of calibrator sources.

Some objects have data from the JCMT Gould Belt Survey (GBS; Ward-Thompson et al. 2007) as published in the variability analysis of (Mairs et al. 2017b). These GBS observations were taken between 2012 and 2014, The flux for objects with a positive crossmatch was calibrated using the conversion factors determined by Mairs et al. (2017b). Following the analysis of Mairs et al. (2017b), we used only the mean flux of the additional GBS epochs.

## Appendix B. Notes on IRAS 18270-0153

In a near-IR spectroscopic survey of Class I YSOs, Connelley et al. (2008); Connelley & Greene (2010) detect a new FUor-like star in a member of a close binary in a small group of young stars around IRAS18270-0153. Two stars in this group are observed as part of the survey, which are identified as IRAS18270-0153(W) and IRAS18270-0153(E) in Connelley et al. (2008) and IRAS18270-0153(1) and IRAS18270-0153(6) in Connelley & Greene (2010). Comparison of the coordinates presented in both papers shows that source (1) is the western star, while source (6) is the eastern YSO.

The classification as a FUor-like YSO in Connelley & Greene (2010) is given due to the strength of the $^{12}$CO absorption, which according to the authors is associated with IRAS18270-0153(W). Since then, several works on eruptive YSOs continue to mention the western source as the FUor-like YSO. However, there are several problems with this designation, and we believe that IRAS18270-0153(E) is the source that shows the FUor-like spectra.

In Table 2 of Connelley & Greene (2010), source (6) is identified as FU Ori type showing a high veiling in its K-band spectrum. The near-IR spectrum of source (6) shows emission of [Fe II] at 1.64 $\mu$m, H$_2$O and $^{12}$CO absorption (see Figure 13 of Connelley & Greene 2010). The identification as a FU Ori-type object in their Table 2 agrees with these properties. Table 4 of Connelley & Greene (2010) shows the equivalent width (EW) of [Fe II] emission as (correctly) associated with source (6). However, there seems to be a mix-up when presenting the EWs of the $^{12}$CO absorption. A value of EW($^{12}$CO) = 3.47 Å is given for source (6), while EW($^{12}$CO) = 32.17 Å is given for source (1). The latter values are not in agreement with the spectra shown in Figure 13 of Connelley & Greene (2010) for these YSOs. The value of EW($^{12}$CO)= 32.17Å is also given for the western source in the table showing the FU Ori-type objects found in the survey (see Table 6 in Connelley & Greene 2010). Finally, when commenting on individual sources in appendix B,Connelley & Greene (2010) identify source (6) as FUor-like.

The spectroscopic observations of known FUor and FUor-like YSOs by Connelley & Reipurth (2018) include IRAS18270-0153W. However the coordinates given for the YSO ($\alpha$ = 18:29:38.9, $\delta$ = $-$01:51:06) correspond to the coordinates of the eastern source in Connelley et al. (2008). The detection of [Fe II] emission in the spectrum presented by Connelley & Reipurth (2018) confirms that the eastern source was observed.

The JCMT transient survey (Lee et al. 2021; Mairs et al. 2024) detected variability at 850 $\mu$m close to the coordinates of IRAS18270-0153. Since the centroid of the sub-mm detections were beyond 15" from the coordinates of IRAS18270-0153W, Lee et al. (2021) concluded that the sub-mm variability was not associated with a FUor outburst. However, inspection of the sub-mm maps presented in Johnstone et al. (2018) show that the 850 $\mu$m emission is closer to the location of IRAS18270-0153(E).

In the catalogue we identify IRAS18270-0153(E) as the correct eruptive YSO, and is associated with the 850 $\mu$m variability observed by the JCMT transient survey (Lee et al. 2021; Mairs et al. 2024).

## Appendix C. Table References

In this section, we present the table that contains the reference numbers for the information provided in tables across this work.

## Appendix D. The Website

The information for the sources in the catalogue will be stored and maintained by the star formation group at Seoul National University.[3] The website will be updated as new objects are discovered. The website will also contain a function that will allow users to provide information on new discoveries, and/or provide comments about sources in the catalogue.

Users will be able to download fits files that contain the information presented in this paper. In addition we will provide csv files with the photometry collected for each source (see Appendix A). The information was collected from several publications and surveys that are referred to in the files. We encourage the users to check these publications before using the photometry provided by the catalogue.

---

[3] http://starformation.synology.me:5002/OYCAT/main.html

**Table C.1.** Reference list for Tables 2 to 7

| #   | Reference                        | #   | Reference                        | #   | Reference                        |
| --- | -------------------------------- | --- | -------------------------------- | --- | -------------------------------- |
| 1   | Wachmann (1954)                  | 51  | Aspin et al. (2006)              | 101 | Chen et al. (2013)               |
| 2   | Herbig (1966)                    | 52  | Movsessian et al. (2006)         | 102 | Dunham et al. (2013)             |
| 3   | Welin (1971a)                    | 53  | Ojha et al. (2006)               | 103 | Hillenbrand et al. (2013)        |
| 4   | Welin (1971b)                    | 54  | Skrutskie et al. (2006)          | 104 | Magakian et al. (2013)           |
| 5   | Strom et al. (1976)              | 55  | Tapia et al. (2006)              | 105 | McMahon et al. (2013)            |
| 6   | Gyul'Budagyan & Sarkisyan (1977) | 56  | Acosta-Pulido et al. (2007)      | 106 | Muzerolle et al. (2013)          |
| 7   | Herbig (1977)                    | 57  | Fedele et al. (2007)             | 107 | Ninan et al. (2013)              |
| 8   | Cohen (1978)                     | 58  | Kóspál et al. (2007)             | 108 | Scholz et al. (2013)             |
| 9   | Elias (1978)                     | 59  | Lawrence et al. (2007)           | 109 | Semkov et al. (2013)             |
| 10  | Wenzel (1978)                    | 60  | Lorenzetti et al. (2007)         | 110 | Stutz et al. (2013)              |
| 11  | Chavarria-K. (1979)              | 61  | Persi et al. (2007)              | 111 | Antoniucci et al. (2014b)        |
| 12  | Wenzel (1980)                    | 62  | Reipurth et al. (2007)           | 112 | Antoniucci et al. (2014a)        |
| 13  | Cohen & Schwartz (1983)          | 63  | Stecklum et al. (2007)           | 113 | Audard et al. (2014)             |
| 14  | Graham & Frogel (1985)           | 64  | Connelley et al. (2008)          | 114 | Contreras Peña et al. (2014)     |
| 15  | Kolotilov & Petrov (1985)        | 65  | Di Francesco et al. (2008)       | 115 | Csengeri et al. (2014)           |
| 16  | Mundt et al. (1985)              | 66  | Lucas et al. (2008)              | 116 | Gramajo et al. (2014)            |
| 17  | Reipurth (1985a)                 | 67  | Massi et al. (2008)              | 117 | Herczeg & Hillenbrand (2014)     |
| 18  | Reipurth (1985b)                 | 68  | Sicilia-Aguilar et al. (2008)    | 118 | Holoien et al. (2014)            |
| 19  | Reipurth & Bally (1986)          | 69  | Enoch et al. (2009)              | 119 | Ibryamov et al. (2014)           |
| 20  | Goodrich (1987)                  | 70  | Gutermuth et al. (2009)          | 120 | Kóspál et al. (2014)             |
| 21  | Beichman et al. (1988)           | 71  | Lorenzetti et al. (2009)         | 121 | Maehara et al. (2014)            |
| 22  | Hartmann et al. (1989)           | 72  | Peneva et al. (2009)             | 122 | Mainzer et al. (2014)            |
| 23  | Herbig (1989)                    | 73  | Samus (2009)                     | 123 | Mottram et al. (2014)            |
| 24  | Eisloeffel et al. (1991)         | 74  | Wils et al. (2009)               | 124 | Abrahamyan et al. (2015)         |
| 25  | Staude & Neckel (1991)           | 75  | Aspin et al. (2010)              | 125 | Chiang et al. (2015)             |
| 26  | Staude & Neckel (1992)           | 76  | Audard et al. (2010)             | 126 | Contreras Peña (2015)            |
| 27  | Kenyon et al. (1993)             | 77  | Connelley & Greene (2010)        | 127 | Gutermuth & Heyer (2015)         |
| 28  | Strom & Strom (1993)             | 78  | Leoni et al. (2010)              | 128 | Hackstein et al. (2015)          |
| 29  | Hartmann & Kenyon (1996)         | 79  | Munari et al. (2010)             | 129 | Hodapp & Chini (2015)            |
| 30  | Hodapp et al. (1996)             | 80  | Peneva et al. (2010)             | 130 | Miller et al. (2015)             |
| 31  | Lada et al. (1996)               | 81  | Semkov et al. (2010)             | 131 | Ninan et al. (2015)              |
| 32  | Reipurth & Aspin (1997)          | 82  | Aspin (2011)                     | 132 | Onozato et al. (2015)            |
| 33  | Sandell & Aspin (1998)           | 83  | Caratti o Garatti et al. (2011)  | 133 | Safron et al. (2015)             |
| 34  | Herbst & Shevchenko (1999)       | 84  | Covey et al. (2011)              | 134 | Tapia et al. (2015)              |
| 35  | Hodapp (1999)                    | 85  | Green et al. (2011)              | 135 | Urquhart et al. (2014)           |
| 36  | Aspin & Sandell (2001)           | 86  | Kóspál et al. (2011b)            | 136 | Benjamin et al. (2016)           |
| 37  | Sandell & Weintraub (2001)       | 87  | Kóspál et al. (2011a)            | 137 | Chambers et al. (2016)           |
| 38  | Prato et al. (2002)              | 88  | Kun et al. (2011)                | 138 | Dodin et al. (2016)              |
| 39  | Reipurth et al. (2002)           | 89  | Lorenzetti et al. (2011)         | 139 | Furlan et al. (2016)             |
| 40  | Aspin & Reipurth (2003)          | 90  | Miller et al. (2011)             | 140 | Jurdana-Šepić & Munari (2016)    |
| 41  | Benjamin et al. (2003)           | 91  | Thommes et al. (2011)            | 141 | Kóspál et al. (2016)             |
| 42  | Hartigan & Kenyon (2003)         | 92  | Alfonso-Garzón et al. (2012)     | 142 | Mairs et al. (2016)              |
| 43  | Briceño et al. (2004)            | 93  | Fischer et al. (2012)            | 143 | Meingast et al. (2016)           |
| 44  | Coffey et al. (2004)             | 94  | Hodapp et al. (2012)             | 144 | Molinari et al. (2016)           |
| 45  | McNeil et al. (2004)             | 95  | Kristensen et al. (2012)         | 145 | Suresh et al. (2016)             |
| 46  | Muzerolle et al. (2004)          | 96  | Lorenzetti et al. (2012)         | 146 | Bonnefoy et al. (2017)           |
| 47  | Persson (2004)                   | 97  | Megeath et al. (2012)            | 147 | Caratti o Garatti et al. (2017)  |
| 48  | Reipurth & Aspin (2004)          | 98  | Reipurth et al. (2012)           | 148 | Contreras Peña et al. (2017b)    |
| 49  | Semkov (2004)                    | 99  | Semkov & Peneva (2012)           | 149 | Contreras Peña et al. (2017a)    |
| 50  | Ojha et al. (2005)               | 100 | Caratti o Garatti et al. (2013)  | 150 | Elia et al. (2017)               |

**Table C.1.** Continued

| # | Reference | # | Reference | # | Reference |
|---|---|---|---|---|---|
| 151 | Fehér et al. (2017) | 201 | Lee et al. (2020a) | 251 | Gaia Collaboration et al. (2023) |
| 152 | Giannini et al. (2017) | 202 | Liu et al. (2020) | 252 | Hillenbrand et al. (2023) |
| 153 | Herczeg et al. (2017) | 203 | Lucas et al. (2020) | 253 | Magakian et al. (2023) |
| 154 | Hunter et al. (2017) | 204 | Sicilia-Aguilar et al. (2020) | 254 | Nagy et al. (2023) |
| 155 | Jurdana-Šepić et al. (2017) | 205 | Siwak et al. (2020) | 255 | Nikoghosyan et al. (2023) |
| 156 | Lucas et al. (2017) | 206 | Stecklum (2020) | 256 | Pokhrel et al. (2023) |
| 157 | Mercimek et al. (2017) | 207 | Szegedi-Elek et al. (2020) | 257 | Singh et al. (2023) |
| 158 | Nikoghosyan et al. (2017) | 208 | Tobin et al. (2020) | 258 | Siwak et al. (2023) |
| 159 | Ruíz-Rodríguez et al. (2017) | 209 | Andreasyan (2021) | 259 | Szabó et al. (2023) |
| 160 | Sicilia-Aguilar et al. (2017) | 210 | Andreasyan et al. (2021) | 260 | Vioque et al. (2023) |
| 161 | Yoo et al. (2017) | 211 | Bailer-Jones et al. (2021) | 261 | Wang et al. (2023) |
| 162 | Ansdell et al. (2018) | 212 | Chen et al. (2021) | 262 | Wang et al. (2023) |
| 163 | Bellm et al. (2018) | 213 | Cutri et al. (2021) | 263 | Ashraf et al. (2024) |
| 164 | Brogan et al. (2018) | 214 | Fiorellino et al. (2021) | 264 | Contreras Peña et al. (2024) |
| 165 | Cieza et al. (2018) | 215 | Guo et al. (2021) | 265 | De Simone et al. (2024) |
| 166 | Connelley & Reipurth (2018) | 216 | Guzmán-Díaz et al. (2021) | 266 | Fiorellino et al. (2024) |
| 167 | Großschedl et al. (2018) | 217 | Hillenbrand et al. (2021) | 267 | Giannini et al. (2024) |
| 168 | Hillenbrand et al. (2018) | 218 | Hillenbrand (2021) | 268 | Guo et al. (2024a) |
| 169 | Jayasinghe et al. (2018) | 219 | Hodgkin et al. (2021) | 269 | Guo et al. (2024b) |
| 170 | Liu et al. (2018) | 220 | Hunter et al. (2021) | 270 | Kim et al. (2024) |
| 171 | Szymczak et al. (2018) | 221 | Kóspál et al. (2021) | 271 | Kuhn et al. (2024) |
| 172 | Arun et al. (2019) | 222 | Kuhn et al. (2021) | 272 | Le Gouellec et al. (2024) |
| 173 | Bøgelund et al. (2019) | 223 | Lee et al. (2021) | 273 | Lucas et al. (2024) |
| 174 | Brogan et al. (2019) | 224 | McMahon et al. (2021) | 274 | Mairs et al. (2024) |
| 175 | Contreras Peña et al. (2019) | 225 | Mège et al. (2021) | 275 | Marton et al. (2024) |
| 176 | Eden et al. (2019) | 226 | Park et al. (2021b) | 276 | Nayakshin et al. (2024) |
| 177 | Hodapp et al. (2019) | 227 | Park et al. (2021a) | 277 | Temmink et al. (2024) |
| 178 | Kun et al. (2019) | 228 | Semkov et al. (2021) | 278 | Tran et al. (2024) |
| 179 | Magakian et al. (2019) | 229 | Stecklum et al. (2021) | 279 | Wolf et al. (2024) |
| 180 | Postel et al. (2019) | 230 | Szabó et al. (2021) | 280 | Zhou et al. (2024) |
| 181 | Proven-Adzri et al. (2019) | 231 | Yoon et al. (2021) | 281 | Carvalho & Hillenbrand (2025) |
| 182 | Zsidi et al. (2019) | 232 | Bayandina et al. (2022) | 282 | Contreras Peña et al. (2025) |
| 183 | Zucker et al. (2019) | 233 | Cruz-Sáenz de Miera et al. (2022) | 283 | Frostig et al. (2025) |
| 184 | Andreasyan et al. (2020) | 234 | Diaz-Rodriguez et al. (2022) | 284 | Hillenbrand et al. (2025) |
| 185 | Baek et al. (2020) | 235 | Ghosh et al. (2022) | 285 | Morris et al. (2025) |
| 186 | Borisov & Denisenko (2020) | 236 | Giannini et al. (2022) | 286 | Nagy et al. (2025) |
| 187 | Burns et al. (2020) | 237 | Guo et al. (2022) | 287 | Neha & Sharma (2025) |
| 188 | Chen et al. (2020a) | 238 | Hillenbrand et al. (2022) | 288 | Sheehan et al. (2025) |
| 189 | Chen et al. (2020b) | 239 | Hodapp et al. (2022) | 289 | Siwak et al. (2025) |
| 190 | Cheng et al. (2020) | 240 | Kuhn et al. (2022) | 290 | Smith et al. (2025) |
| 191 | Connelley & Reipurth (2020) | 241 | Mutafov et al. (2022) | | |
| 192 | Contreras Peña et al. (2020) | 242 | Park et al. (2022) | | |
| 193 | Dahm & Hillenbrand (2020) | 243 | Szabó et al. (2022) | | |
| 194 | Dutta et al. (2020) | 244 | Valdivia-Mena et al. (2022) | | |
| 195 | Guo et al. (2020) | 245 | Yoon et al. (2022) | | |
| 196 | Hankins et al. (2020) | 246 | Zakri et al. (2022) | | |
| 197 | Hillenbrand et al. (2020) | 247 | Burns et al. (2023) | | |
| 198 | Hodapp et al. (2020) | 248 | Carvalho et al. (2023) | | |
| 199 | Kóspál et al. (2020) | 249 | Contreras Peña et al. (2023a) | | |
| 200 | Lee et al. (2020b) | 250 | Contreras Peña et al. (2023b) | | |